\documentclass{article}
\usepackage[dvips]{graphicx}
\begin{document}

\title{Time dependence of cosmological and inflationary parameters in slow-roll inflation}

\author{Shiro Hirai and Tomoyuki Takami\\
\\
Department of Digital Games, Osaka Electro-Communication University\\
1130-70 Kiyotaki, Shijonawate, Osaka 575-0063, Japan
}

\maketitle

\textbf{\Large Abstract} \\ \\
The dependence of cosmological and inflationary parameters on time during the last 60 $e$-folds in inflation is investigated using a slow-roll inflation model. The time dependence of the inflaton field is calculated for the case of chaotic inflation by a numerical method rather than the familiar approximations. It is found that the Hubble parameter and the spectral index decrease in the last 60 $e$-folds. The dependence of the power spectrum of the curvature perturbation on the size perturbation $k$ is calculated numerically, and it is shown that the overall constant value of the power spectrum thus determined differs from that given by the familiar method. However, the $k$-dependent spectra of the numerical calculation, the Bessel function approximation, and the familiar Taylor expansion are all found to be consistent. \\

PACS number:98.80Cq

\section{Introduction}
Inflation is an important concept in cosmology, and is supported by recent satellite-based measurements. The present authors have studied the effect of the length of inflation and pre-inflation on the power spectrum and the angular power spectrum [1], and it has been shown that the suppression of the spectrum at $l=2$ as indicated by Wilkinson Microwave Anisotropy Probe (WMAP) data [2] may be explained to a certain extent by the finite length of inflation for inflation of 50-60 $e$-folds [3]. The length of inflation is generally believed to be very long, but the physically interesting period of inflation is that in the last 60 $e$-folds. Although a correct inflationary potential based on superstring or supergravity theory has yet to be established, it could be assumed that the potential term in this latter period of inflation can be expressed in a simple form, such as a small-field or large-field model, or a hybrid model. In the present paper, the time-dependent behavior of the inflaton field during the last 100 $e$-folds in inflation is investigated using a simple inflation model. Specifically, numerical calculations are performed using a chaotic inflation model ($\phi^2$ model), and the effects of the time dependence of the inflaton field on cosmological and inflationary parameters such as slow-roll parameters, the Hubble parameter, and the spectral index are investigated. The $k$-dependence of the power spectrum of the curvature perturbation has been investigated using Taylor expansion or the numerical method [4-7]. Here, using the derived parameters, i.e., the slow-rolled parameters, Hubble parameter and  the inflaton field  which have the time dependence, the power spectrum is calculated with the numerical method and the Bessel approximation, and it is compared with the power spectra derived from the familiar methods.

\section{Cosmological and inflationary parameters}

The dependence of the cosmological and inflationary parameters on time is investigated using a slow-roll inflation model. Assuming a spatially flat universe and the energy density of the universe to be dominated by the inflaton field, the Einstein field equations can be written as 
\begin{equation}
H^2 = \frac{8\pi}{3m^2}(V(\phi)+\frac{1}{2}\dot{\phi}^2),
\end{equation}
and
\begin{equation}
\ddot{\phi}+3H\dot{\phi}+V^\prime(\phi)=0,
\end{equation}
where $\phi$ is the inflaton field, $V(\phi)$ is the inflaton potential, $m$ is the Planck mass, and $H=\dot{a}/a$ with $a$ being a scale factor. The variable $N$ defined by $dN = Hdt$ is introduced to represent time with respect to the number of $e$-folds of inflation. Equation (1) can then be written using the relation $\dot{\phi}=Hd\phi/dN$ as
\begin{equation}
H^2=\frac{8\pi V(\phi)}{3m^2(1-4\pi/3m^2(d\phi/dN)^2)}.
\end{equation}
The Hubble parameter is usually considered to be a constant, i.e., $H^2 \cong 8\pi/3m^2V(\phi)$ . In the generalization, however, with $\phi$ being dependent on time, the Hubble parameter also includes a time dependence (from eq. (3)). Using the variable $N$, eq. (2) can be written as
\begin{equation}
\frac{d^2\phi}{dN^2}+\frac{1}{H}\frac{dH}{dN}\frac{d\phi}{dN}+3\frac{d\phi}{dN}+\frac{V^\prime(\phi)}{H^2}=0.
\end{equation}
Equation (4) can be further written in terms of $\phi$ and the derivate of $\phi$ with respect to $N$. To solve eq. (4) numerically, the slow-roll inflation model is adopted, which is a chaotic inflation model with potential term given by $V(\phi)=M^4/2(\phi/m)^2$ . Using an initial condition of $N = n_0$, 
\begin{equation}
\phi(n_0)=\frac{\sqrt{-n_0}}{\sqrt{2\pi}}m ,
\end{equation}
\begin{equation}
\frac{d\phi(n_0)}{dN}=-\frac{m}{\sqrt{8\pi(-n_0)}} .
\end{equation}
The numerical solutions for eq. (4) for $n_0$ values of $-100 $ to $-60 $ do not reveal any appreciable dependence on $n_0$. The behavior of $\phi$ in the case of $n_0=-100$ is shown in Figure 1. The inflaton field can be seen to slowly decrease. The behavior of the Hubble parameter according to eq. (3), which includes $\phi$, is shown in Figure 2. The Hubble parameter becomes considerably smaller as $N$ approaches zero. The value $H^2(N)/H^2(-60)$ is 1.33 for $N = -80$, 0.835 for $N = -50$, 0.669 for $N = -40$, and 0.504 for $N = -30$.
   For slow-roll inflation, the following parameters are employed [4]:
\begin{equation}
\epsilon=3\frac{\dot{\phi}^2}{2}(\frac{\dot{\phi}^2}{2}+V)^{-1}=\frac{m^2}{4\pi}(\frac{H^\prime(\phi)}{H(\phi)})^2,
\end{equation}
\begin{equation}
\delta=\frac{m^2}{4\pi}\frac{H^{\prime\prime}(\phi)}{H(\phi)},
\end{equation}
\begin{equation}
\xi=\frac{m^4}{16\pi^2}\frac{H^{\prime}(\phi)H^{\prime\prime\prime}(\phi)}{(H(\phi))^2}.
\end{equation}
Other slow-roll parameters ($\epsilon_V$, $\eta_V$, $\xi_V$) can be written in terms of the slow-roll parameters $\epsilon$, $\delta$, and $\xi$ to the first order in slow roll: $\epsilon=\epsilon_V$, $ \delta=\eta_V-\epsilon_V$, and $\xi=\xi_V-3\epsilon_V\eta_V+3\epsilon^2_V$, where  $\epsilon_V=m^2/16\pi(V^\prime/V)^2$, $\eta_V=m^2/8\pi(V^{\prime\prime}/V)$, and $\xi=m^4/64\pi^2(V^\prime V^{\prime\prime\prime}/V^2)$. Figure 3 shows the behavior of $\epsilon$. The value of $\epsilon$ can be seen to remain small ($\epsilon(-60)=0.0083$, $\epsilon(-50)=0.0099$, and $\epsilon(-40)=0.0124$) for $N$ smaller than $-10$. The value of the important parameter $\delta$ is zero in the first-order approximation using this $\phi^2$ model, and the values of $\delta$ and $\xi$ are very small. 
The time dependence of the spectral index $n_s$ can also be calculated in this manner. The spectral index is given by 
\begin{equation}
n_s=1+2\delta-4\epsilon
\end{equation}
and the time dependence of $n_s$ is shown in Figure 4. The value of $n_s$ decreases as $N$ approaches zero: $n_s(-80)=0.981$, $n_s(-60)=0.975$, $n_s(-50)=0.970$, and $n_s(-40)=0.963$.

\section{Scalar perturbations}

The dependence of the curvature perturbation on the parameter $k$ is investigated assuming a spatially flat Friedman-Robertson-Walker (FRW) universe with a background spectrum given by metric perturbations. The line element for the background and perturbations is generally expressed as [8] 
\begin{equation}
ds^2=a^2(\tau)\{ (1+2A)d\tau^2 -2\partial_i Bdx^i d\tau-[(1-2\Psi)\delta_{ij}+2\partial_i \partial_j E+h_{ij}] dx^i dx^j \},
\end{equation}
where $\tau$ is the conformal time, the functions $A$, $B$, $\Psi$, and $E$ represent the scalar perturbations, and $h_{ij}$ represents tensor perturbations. The density perturbations in terms of the intrinsic curvature perturbation of comoving hypersurfaces is given by $\mathcal{R}=-\Psi-(H/\dot{\phi})\delta\phi$, where $\phi$ is the inflaton field, $d\phi$ is the fluctuation of the inflaton field, and $\mathcal{R}$ is the curvature perturbation. Overdots represent derivatives with respect to time $t$, and primes represent derivatives with respect to the conformal time $\tau$. Introducing the gauge-invariant potential $u\equiv a(\tau)(\delta\phi+(\dot{\phi}/H)\Psi)$ allows the action for scalar perturbations to be written as [9]
\begin{equation}
S=\frac{1}{2}\int d\tau d^3x\{ (\frac{\partial u}{\partial \tau})^2-(\nabla u)^2+\frac{Z^{\prime\prime}}{Z}u^2 \},
\end{equation}
where $Z=a\dot{\phi}/H$ and $u=-Z\mathcal{R}$. 
The field $u(\tau, x)$ is expressed using annihilation and creation operators as 
\begin{equation}
u(\tau , \mathbf{x})=\frac{1}{(2\pi)^{3/2}}\int d^3k\{ u_k(\tau)a_\mathbf{k}+u_k^\ast(\tau)a_{-\mathbf{k}}^\dagger \}e^{-i\mathbf{kx}},
\end{equation}
and the field equation for $u_k(\tau)$ is derived as
\begin{equation}
\frac{d^2u_k}{d\tau^2}+(k^2-\frac{1}{Z}\frac{d^2Z}{d\tau^2})u_k=0,
\end{equation}
where the solution to $u_k$ satisfies the normalization condition $u_kdu_k^\ast/d\tau-u_k^\ast du_k/d\tau=i$. 
Using the slow-roll parameters, $(d^2Z/d\tau^2)/Z$ is written exactly as
\begin{equation}
\frac{1}{Z}\frac{d^2Z}{d\tau^2}=2a^2H^2\nu,
\end{equation}
where
\begin{equation}
\nu=(1+\epsilon-\frac{3}{2}\delta+\epsilon^2-2\epsilon\delta+\frac{\delta^2}{2}+\frac{\xi}{2}).
\end{equation}
A partial integration allows $\tau$ to be expanded as follows [10].
\begin{equation}
\tau=-\frac{1}{aH(1-\epsilon)}-\frac{2\epsilon(\epsilon-\delta)}{aH(1-\epsilon)}+\int\frac{\epsilon(-2\delta^2+\delta(9-4\epsilon)\epsilon-6\epsilon^2+3\epsilon^3-\xi(1-\epsilon))}{(1-\epsilon)^2}\frac{dN}{aH}.
\end{equation}
Here, using the approximate relation $a^2H^2=(1+2\epsilon(\epsilon-\delta))^2/\tau^2(1-\epsilon)^2$ , eq. (14) can be rewritten as 
\begin{equation}
\frac{d^2u_k}{d\tau^2}+(k^2-\frac{2\nu(1+2\epsilon(\epsilon-\delta))^2}{\tau^2(1-\epsilon)^2})u_k=0.
\end{equation}
If it is assumed that the slow-roll parameters can be constant, the solution for eq. (18) can be written using the Hankel function $H^{(1)}_\mu(-k\tau)$ as
\begin{equation}
u_k(\tau)=\frac{\sqrt{\pi}}{2\sqrt{k}}e^{i(\mu+1/2)\pi/2}(-k\tau)^{1/2}H^{(1)}_\mu(-k\tau),
\end{equation}
where $\mu=\sqrt{2\nu(1+2\epsilon(\epsilon-\delta))^2/(1-\epsilon)^2+1/4}$. This solution is fixed such that as $k\tau\to -\infty $, $u_k(\tau)$ approach plane waves. The numerical solution to eq. (14) can then be considered. Equation (14) can be written in terms of $N$ as 
\begin{equation}
\frac{d^2u_k}{dN^2}+\frac{1}{aH}\frac{daH}{dN}\frac{du_k}{dN}+(\frac{k^2}{(aH)^2}-2\nu)u_k=0.
\end{equation}
In order to calculate eq. (20) numerically, it is assumed that the present-day size perturbation $k=0.002$(1/Mpc) exceeds the Hubble radius in inflation at the time of $N=-60$. The scale factor is thus written as $a=0.002/H_0e^{N+60}$, where $H_0$ is the value of $H$ at $N=-60$. As the Hubble parameter and $\nu$ have both been calculated as functions of $N$, eq. (20) can be calculated numerically with an initial value of $u_k$ derived from eq. (19). Note that the desired solution of $u_k$ is that at the point $aH=k$. In Figure 5, the absolute values of $u_k$ derived by numerical calculation of eq. (20) at $aH=k$ is shown as a function of k. The absolute value of $u_k$ (eq.(19)) obtained using the Bessel approximation, adopting the value of $\mu$ as the numerically calculated value, is also shown. The value of $\mu$ changes slightly according to $N$: $\mu(-80)=1.513$, $\mu(-60)=1.517$, $\mu(-50)=1.520$, $\mu(-40)=1.525$, and $\mu(-30)=1.534$. The results shown in Figure 5 do not suggest any appreciable difference between the numerical calculation and the Bessel approximations, indicating that the Bessel approximation, in which $\mu$ exhibits $k$ dependence, can be considered satisfactory.\\
\hspace*{1pc} The power spectrum of the scalar perturbations $P_\mathcal{R}$ is defined as follows [4].
\begin{equation}
<\mathcal{R}_\mathbf{k}(\eta), \mathcal{R}_\mathbf{l}^\ast(\eta)> = \frac{2\pi^2}{k^3}P_\mathcal{R}\delta^3(\mathbf{k}-\mathbf{l}).
\end{equation}
Here, $\mathcal{R}_k(\eta)$ is the Fourier series of the curvature perturbation $\mathcal{R}$. The power spectrum $P_\mathcal{R}$ is then written as [4]
\begin{equation}
P_\mathcal{R} = (\frac{k^3}{2\pi^2})\bigl|\frac{u_k}{Z}\bigr|^2,
\end{equation}
where $Z=a\dot{\phi}/H$. The power spectrum is estimated by four different methods; the numerical method, the Bessel approximation using eq. (19), and Taylor expansion with and without running as two familiar methods. The numerical calculation of the power spectrum is performed by the present method for $k=aH$. In the case of the Bessel approximation using the values of $\mu$, $d\phi/dN$, and $H$ derived from the numerical calculation (having $k$ dependence), the power spectrum (22) can thus be written as 
\begin{equation}
P_\mathcal{R} = \frac{k^2}{8\pi}\frac{(-k\tau)(H^{(1)}_\mu(-k\tau))^2}{a^2(d\phi/dN)^2}|_{k=aH},
\end{equation}
where the value of $P_\mathcal{R}$ is obtained at $k=aH$ (i.e. $k\tau=-(1+2\epsilon(\epsilon-\eta))/(1-\epsilon))$). There is very little difference between the results of the numerical calculation and the Bessel approximation (see Table 1), again demonstrating that the Bessel approximation is satisfactory. 
In the first of the familiar cases, the power spectrum is written as 
\begin{equation}
P_\mathcal{R} = \nu^{2\mu-1}(\frac{\Gamma(\mu)}{\Gamma(3/2)})^2(\mu-1/2)^{1-2\mu}\frac{H^4}{m^4|H^\prime|^2}|_{k=aH}.
\end{equation}
In the alternate case, the power spectrum is given by [5]
\begin{equation}
P_\mathcal{R} = (1-(2C+1)\epsilon+C\delta)^2\frac{4H^4}{m^4|H^\prime|^2}|_{k=aH},
\end{equation}
where $C=-2+\ln 2+\gamma\cong -0.73$, with $\gamma$ being the Euler constant. The value of the power spectrum ($P_\mathcal{R}$) given by eq. (24) is $3049.07(M^4/m^4)$, and that by eq. (25) is $3100.34(M^4/m^4)$ assuming the value of $\mu$ at $N=-60$. Using the same parameter values, the numerical calculation by eq. (22) and the Bessel approximation (eq. (23)) both afford a value of $6252.43(M^4/m^4)$. There is thus a difference of approximately a factor of two between the present numerical and Bessel calculations and the familiar cases (eqs.(24) and (25)). This discrepancy can be attributed to the difference in the Hankel function used for deriving the expressions. The familiar expressions are derived from the asymptotic form ($k/aH\to 0$) of the Hankel function, and are estimated at $k/aH=1$[4], whereas the present Bessel calculation uses the correct form. The difference between the asymptotic and correct forms of the Hankel function equates to a factor of approximately $\sqrt{2}$ at $k/aH=1$, giving rise to a change in the normalization of the power spectrum.\\    
\hspace*{1pc} As the value of the spectral index differs from unity, as indicated by experimental results such as the WMAP data [2, 11], the power spectrum can be inferred to have $k$ dependence. The power spectrum is usually expanded at a pivot scale $k_0$, i.e. [6],
\begin{equation}
P_\mathcal{R}(k) = P_\mathcal{R}(k_0)(\frac{k}{k_0})^{n_s(k_0)-1+(1/2)\alpha \ln(k/k_0)},
\end{equation}
where $\alpha=dn_s/d\ln k|_{k=k_0}$. Two cases are considered here; $\alpha=0$ (no running of the spectral index), and $\alpha\neq 0$ (running). Table 1 lists the results for the four calculations at various values of $k$, and Figure 6 compares the numerically calculated power spectrum with that determined by Taylor expansion with power spectrum running, where the over-all factor $ P_\mathcal{R}(k_0) $ is fixed as $6252.43 (M^4/m^4)$. These results confirm that the power spectrum given by Taylor expansion with running exhibits very similar behavior to that obtained by the present numerical calculation, and that there is very little difference among the power spectrum values for the numerical, Bessel, and Taylor expansion with running cases. The largest difference, 0.05\% occurs at $k = 1 $ (1/Mpc), while the Taylor expansion without spectral running differs by up to 1\%. 

\section{Discussion and summary}
The time dependence of the cosmological and inflationary parameters and the $k$ dependence of the power spectrum of the curvature perturbation during  the last 100 $e$-folds in inflation were investigated numerically using a slow-roll inflation model. Using a time dependence for the inflaton field calculated for the case of chaotic inflation ($\phi^2$ model), it was shown that both the Hubble parameter and spectral index decrease with time (see Figures 2 and 4). The time dependence of the slow-roll parameters was also derived. It was shown that an approximation of the Hankel function obtained by changing the slow-roll parameters is closely consistent with the numerical calculations, and that there exists an approximately two-fold difference in normalization between the present methods and the usual methods due to the asymptotic form of the Hankel function in the usual treatments. In the $ k $-dependent spectra, the numerical and Bessel calculations are very similar to the Taylor expansion with spectral running, although the Taylor expansion without running differs by up to 1\% at $k = 1 $ (1/Mpc). These results indicate that the running of the spectral index should be considered in the derivation of an inflation model. Although calculations were performed in the present study for chaotic inflation with potential of $\phi^2$, this approach can be similarly applied to any arbitrary potential of inflation. It is inferred that similar behavior as derived in the present case will be obtained for other slow-roll models. Detailed analyses using other slow-roll models will be performed as part of future work.\\ \\

\textbf{\large Acknowledgments}\\[0.1cm]
The authors would like to thank the staff of Osaka Electro-Communication University for valuable discussions.\\[0.1cm]

\textbf{\large References}\\[0.1cm]
[1] Hirai S 2003 {\it Class.Quantum Grav.} \textbf{20} 1673 (hep-th/0212040); Hirai S 2003 (hep-th/0307237); Hirai S 2005 {\it Class.Quantum Grav.} \textbf{22} 1239 (astro-ph/0404519)\\[0.1mm]
[2] Bennett C L et al. 2003 {\it Astrophys.J., Suppl. Ser.} \textbf{146} 1 (astro-ph/0302207); Spergel D N et al. 2003 {\it Astrophys. J., Suppl. Ser.} \textbf{146}, 175 (astro-ph/0302209); Peiris H V et al. 2003 {\it Astrophys. J., Suppl. Ser.} \textbf{146} 213 (astro-ph/030225)\\[0.1mm] 
[3] Hirai S and Takami T 2006 {\it Class.Quantum Grav.} \textbf{23} 2541 (astro-ph/0506479)\\[0.1mm]
[4] Lidsey J E et al. 1997 {\it Rev.Mod.Phys.} \textbf{69} 373\\[0.1mm]
[5] Schwarz D J, Cesar A, Terrero-Escalate and Garcia A A 2001 {\it Phys.Lett. B} \textbf{517} 243\\[0.1mm]
[6] Peiris H V et al. 2003 {\it Astrophys. J. Suppl.} \textbf{148} 213\\[0.1mm]
[7] Hamann J, Lesgourgues J and Valkenburg W (astro-ph/0802.0505); 
Ringeval C  2008 {\it Lect.NotesPhys.} 738:243 (astro-ph/0703486); 
Martin J and Ringeval C 2006 {\it JCAP} 0608 009 (astro-ph/0605367)\\[0.1mm]
[8] Bardeen J M 1980 {\it Phys.Rev.} \textbf{D22} 1882; Kodama H and Sasaki M 1984 {\it Prog.Theor.Phys.Suppl.} \textbf{78} 1\\[0.1mm]
[9] Mukhanov V F, Feldman H A and Brandenberger R H 1992 {\it Phys.Rep.} \textbf{215} 203\\[0.1mm]
[10] Schwarz D J, Cesar A, Terrero-Escalate and Garcia A A 2001 {\it Phys.Lett. B} \textbf{517} 243\\[0.1mm]
[11] Dunkeley J et al. 2008 (WMAP Collaboration) arXir:0803.0586\\[0.1mm]

\begin{table}[h]
  \caption{Power spectrum values ($P_\mathcal{R}[M^4/m^4]$) and ratio to numerical calculation}
  \label{aaaaa}
  \begin{center}
    \begin{tabular}{|c|c|c|c|c|} \hline
$k$(1/Mpc) & Numerical & Bessel	& Taylor(running) & Taylor(no running)\\ \hline
0.002 & 6252.43 & 6252.43 & 6252.43 & 6252.43\\ \hline
0.01 & 5923.03 & 5923.39 & 5921.54 & 5925.80\\ \hline
(ratio) & (1) & (1.00006) & (0.999749) & (1.00047)\\ \hline
0.1 & 5467.71 & 5468.01 & 5464.68 & 5487.96\\ \hline
(ratio) & (1) & (1.00006) & (0.999446) & (1.0037)\\ \hline
1.0 & 5030.45 & 5030.75 & 5028.23 & 5028.23\\ \hline
(ratio) & (1) & (1.00006) & (0.999559) & (1.01034)\\ \hline
    \end{tabular}
  \end{center}
\end{table}

\includegraphics[width=12cm, clip]{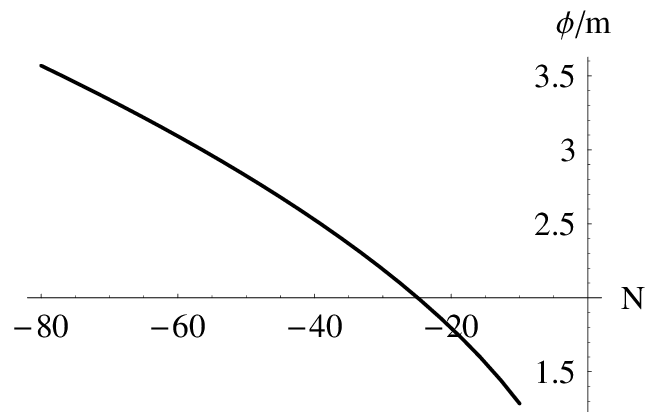}\\
Figure 1: Variation in $\phi/m$ ($m$ is Planck mass) with respect to $N$.\\[0.5cm]

\includegraphics[width=12cm, clip]{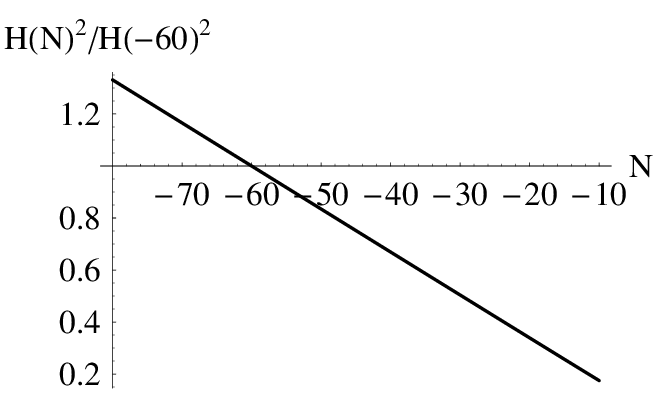}\\
Figure 2: Variation in the Hubble parameter ($H^2(N)/H^2(-60)$) with respect to $N$.\\[0.5cm]

\includegraphics[width=12cm, clip]{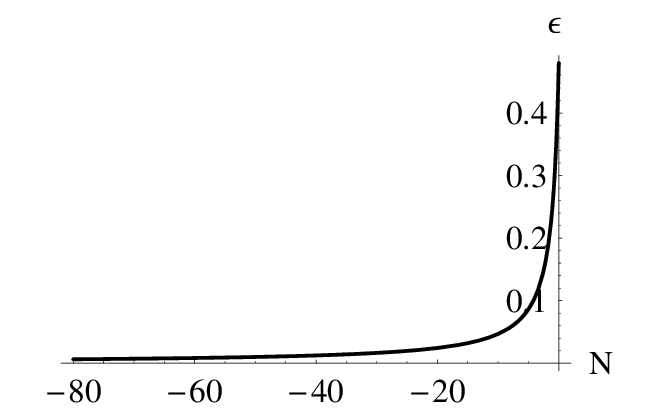}\\
Figure 3: Variation in $\epsilon$ with respect to $N$.\\[0.5cm]

\includegraphics[width=12cm, clip]{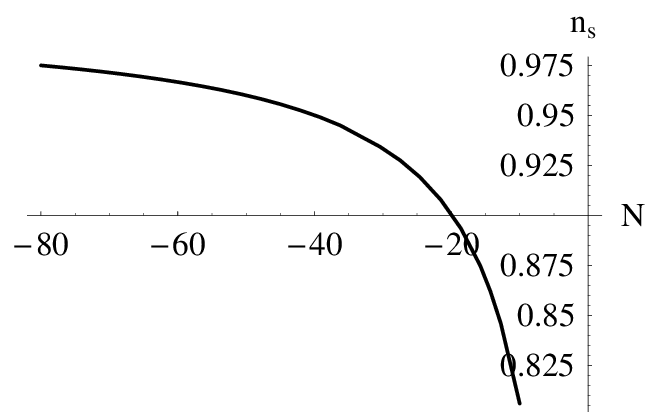}\\
Figure 4: Variation in $n_s$ with respect to $N$.\\[0.5cm]

\includegraphics[width=12cm, clip]{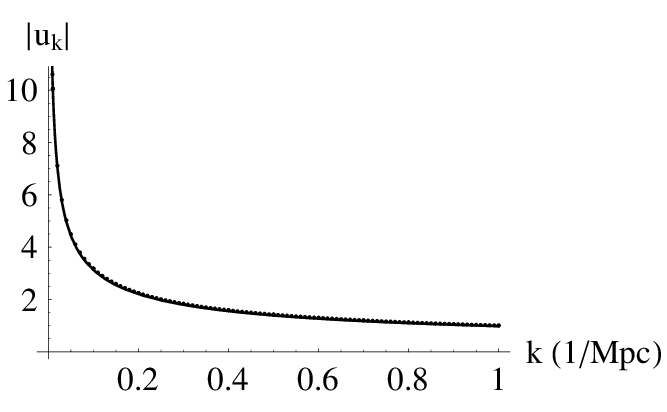}\\
Figure 5: Variation in $u_k$ with respect to $k$. Dotted line denotes the numerical calculation, and solid red line denotes the Bessel approximation using the numerically calculated value of $\mu$.\\[0.5cm]

\includegraphics[width=12cm, clip]{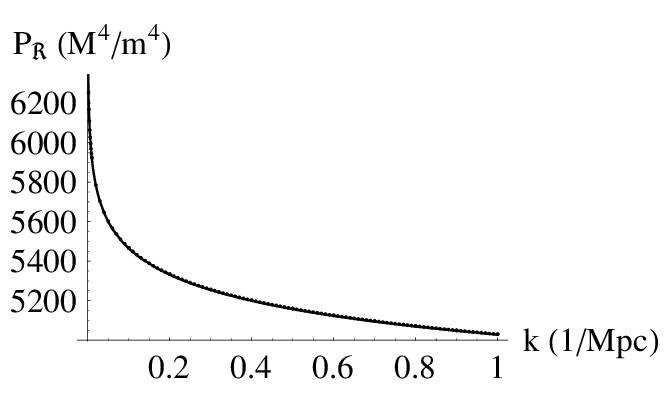}\\
Figure 6: Variation in $P_\mathcal{R}$ with respect to $k$. Dotted line denotes the numerical calculation, and the red solid line denotes the Taylor approximation with spectral index running for the same value of $P_\mathcal{R}(k_0)$.\\[0.5cm]

\end{document}